\documentclass[a4paper,12pt]{article}
\usepackage[pctex32]{graphics}
\usepackage{amssymb,amsmath}

\begin{document}

\topmargin 0pt
\oddsidemargin 0mm
\newcommand{\be}{\begin{equation}}
\newcommand{\ee}{\end{equation}}
\newcommand{\ba}{\begin{eqnarray}}
\newcommand{\ea}{\end{eqnarray}}
\newcommand{\fr}{\frac}

\renewcommand{\thefootnote}{\fnsymbol{footnote}}

\begin{titlepage}

\vspace{5mm}
\begin{center}
{\Large \bf Nonpropagation of massive mode on AdS$_2$ \\
in topologically massive gravity}

\vskip .6cm
 \centerline{\large
 Yong-Wan Kim $^{1,a}$, Yun Soo Myung$^{1,b}$,
and Young-Jai Park$^{2,c}$}

\vskip .6cm

{$^{1}$Institute of Basic Science and School of Computer Aided
Science,
\\Inje University, Gimhae 621-749, Korea \\}

{$^{2}$Department of Physics and Center for Quantum Spacetime, \\
Sogang University, Seoul 121-742, Korea}
\end{center}

\vspace{5mm}
 \centerline{{\bf{Abstract}}}

\vspace{5mm}

Making use of Achucarro-Ortiz (AO) type of dimensional reduction, we
study the topologically massive gravity with a negative
cosmological constant on AdS$_2$ spacetimes. For a constant
dilaton, this two-dimensional model also admits three AdS$_2$
vacuum solutions, which are related to two AdS$_3$ and warped
AdS$_3$ backgrounds with an identification upon uplifting three
dimensions. We carry out the perturbation analysis around these
backgrounds to find what is a physically propagating field.
However, it turns out that there is no propagating massive mode on
AdS$_2$ background, in contrast to the Kaluza-Klein (KK) type of
dimensional reduction. We note  that two dimensionally reduced
actions are different and thus,  the non-equivalence of their
on-shell amplitudes is obtained.

\vspace{5mm}


\vskip 0.8cm

\noindent $^a$ywkim65@gmail.com \\
\noindent $^b$ysmyung@inje.ac.kr \\
\noindent$^c$yjpark@sogang.ac.kr

\noindent
\end{titlepage}

\newpage

\renewcommand{\thefootnote}{\arabic{footnote}}
\setcounter{footnote}{0} \setcounter{page}{2}

\section{Introduction}
The gravitational Chern-Simons terms with a coupling constant $K$
in 3D Einstein gravity produce a physically propagating massive
graviton~\cite{DJT}. This topologically massive gravity with a
negative cosmological constant $\Lambda=-1/l^2$ (TMG$_\Lambda$)
gives us the AdS$_3$ solution~\cite{LSS1,LSS2}. For the Newton's
constant $G_3>0$, the massive modes carry negative energy on the
AdS$_3$ background. In this sense, the AdS$_3$ background may  not
be a stable vacuum for $K\not=0$. The opposite case of $G_3<0$ may
cure the problem, but induces a negative mass for the BTZ black
hole. Another relevant issue is what is the number $N_c$ of
physical degrees of freedom (DOF) for a massive mode for
$K\not=0$. Now one may arrive at  a consensus with $N_c=1$.

There is a possibility for avoiding negative energy when choosing
the chiral point of $ K=l$.  At this point, the massive mode
becomes a massless left-moving graviton, which carries no energy
and may be considered as a pure gauge. However, this special point
raises many
questions~\cite{CDWW1,GJ,GKP,Park,GJJ,CDWW2,Carl,Stro,ssol,MyungL}.
At the chiral point of $K=l$,  a physical degree of freedom seems
to be propagating when using the linearized analysis.
Even though the  canonical analysis has
shown that the  DOF is one ($N_c=1$) at the chiral point~\cite{GJJ,Carl,BC}, a single DOF is
not yet  confirmed  to be $\psi^{new}$ which is regarded as an
additional mode for the graviton~\cite{MyungL}. Actually, as was
claimed in~\cite{BC}, the transition to the chiral coupling does
not make a critical change on the form of the Poisson bracket
algebra. This may imply that a  choice of $K=l$ is nothing special
in  the canonical analysis.

In general, it is not easy to handle the gravitational
Chern-Simons terms  even though they are invariant under
diffeomorphism and Weyl rescaling of the metric~\cite{GIJP,GK}.
Hence, one needs to seek  another way to investigate the
TMG$_\Lambda$. In this end, one may introduce conformal
transformation to single out a conformal degree of freedom (a
dilaton $\phi$) and then use the Kaluza-Klein (KK) ansatz to
obtain an effective two-dimensional action of 2DTMG$_\Lambda$,
which will be a gauge and coordinate invariant. Actually, this was performed
by introducing the metric of $ds^2_{KK}=\phi^2[g_{\mu\nu}dx^\mu
dx^\nu + (d\theta+A_\mu dx^\mu)^2]$. Saboo and Sen~\cite{SSen}
have used the 2DTMG$_\Lambda$ to obtain the entropy of extremal
BTZ black hole~\cite{BTZ} by applying  the entropy function
formalism. Furthermore, the authors in~\cite{AFM} have used the
entropy function approach to find three distinct vacuum solutions
of the 2DTMG$_\Lambda$.

Recently, we have studied the topologically massive gravity with a
negative cosmological constant on AdS$_2$ spacetimes by making use
of the KK type of dimensional reduction \cite{mkp1}. For
a constant dilaton, this two-dimensional model of
2DTMG$_{\Lambda}$ admits three AdS$_2$ vacuum solutions, which are
related to the AdS$_3$ with a positive/negative charge and warped
AdS$_3$ backgrounds \cite{GC1,GC2,StromingerWp,GC3,CD,OK} with an
identification upon uplifting three dimensions. We have carried
out the perturbation analysis around these backgrounds to find
what is a physically propagating field on the AdS$_2$ background
whose curvature is $\bar{R}=-2/v$. It turns out that  a
perturbation mode $\delta F = (h-f/e)$ of $F$ as a dual scalar of
the Maxwell field  is a nonpropagating field in the
absence of the Chern-Simons terms.  However, it becomes a massive
mode in the presence of the Chern-Simons terms, whose equation is
given by $( \bar{\nabla}^2 - m_{\pm}^2 ) \delta F=0 $ with
$m_{\pm}^2 = \frac{2}{v} - \frac{1}{4v}\left(1 \pm
\frac{l}{K}\right) \left(5 \mp \frac{l}{K} \right)$ for two
AdS$_2$ solutions. This confirms
clearly that the DOF  is one ($N_c=1$) for the topologically massive gravity
\footnote{On the other hand, we
expect to have a massive mode with 2 DOF when adding the
Pauli-Fierz action of ${\cal I}_{PF}= m^2_{PF}\int
d^3x\sqrt{-\bar{g}}[h^{mn}h_{mn}-(h_m^m)^2]$~\cite{FP}. However,
we have to use a different method to deal with  a massive field in
the TMG$_{\Lambda}$, which is supposed to be a massive mode
including a gravivector. The reason is clear because the bilinear
Chern-Simons term of ${\cal I}_{CS}=K\int d^3x
\sqrt{-\bar{g}}h^{mn}{\cal C}^{(1)}_{mn}$ is basically different
from ${\cal I}_{PF}$, where ${\cal C}^{(1)}_{mn}$ is the
linearized Cotten tensor~\cite{LSS1}. It was shown that ${\cal
I}_{CS}$ is a square-root of ${\cal I}_{PF}$~\cite{BHT}.}.

In this work, we will consider the  AO type
of dimensional reduction (AOTMG$_\Lambda$) based on the metric of Eq.
(\ref{AOmetric}) introduced  by Achucarro
and Ortiz~\cite{AO}.  They
had exactly recovered the correspondence between the BTZ black
hole and AdS$_2$ black hole. In this approach, we note that the
``linear dilaton" was mainly used for recovering the AdS$_2$ black
hole~\cite{Nav,LMK,GM,mkp2,mkp3,mkp4} and confirming the
AdS$_2$/CFT$_1$ correspondence~\cite{ads21,ads23,ads24}.
In the presence of the gravitational
Chern-Simons terms~\cite{AD}, the authors in~\cite{ads22} have
first considered the AOTMG$_\Lambda$ to find AdS$_2$ solutions.

For ``constant dilation", we will perform perturbation analysis
for the AOTMG$_\Lambda$. This means that we are working  in the
near-horizon geometry AdS$_2 \times S^1$ of  extremal BTZ black
hole.  A perturbation mode $\delta F= (h-f/e)$  is a gauge-artifact in the absence of
the Chern-Simons terms. We find that $\delta F$  remains a
massless, redundant mode  even in the presence of the Chern-Simons
terms, in contrast to the previous result for the KK
reduction case~\cite{mkp1}.

\section{TMG$_\Lambda$ }
Now, let us start with the action for  topologically massive gravity with a negative
cosmological constant (TMG$_\Lambda$) given by~\cite{DJT}
\begin{equation} \label{tmg}
I_{\rm TMG_\Lambda}=\frac{1}{16 \pi G_3}\int
d^3x\sqrt{-g}\Bigg[R_3 -2\Lambda +
\frac{K}{2}\varepsilon^{lmn}\Gamma^p_{~lq}
\Big(\partial_{m}\Gamma^q_{~np}+\frac{2}{3}\Gamma^q_{~mr}\Gamma^r_{~np}\Big)\Bigg],
\end{equation}
where $\varepsilon^{lmn}$ is the tensor  defined by $\epsilon^{lmn}/\sqrt{-g}$
with $\epsilon^{012}=1$. We choose the Newton's constant $G_3>0$.
The Latin indices of $l,m,n, \cdots$ denote three dimensional
tensors. The $K$-term is called the gravitational
Chern-Simons terms.  It is the third order derivative
correction to the 3D Einstein gravity. Here we choose ``+" sign to avoid negative
graviton energy~\cite{GJ}.

Varying this action leads to the Einstein equation
\begin{equation} \label{eineq}
G_{mn} + KC_{mn}=0,
\end{equation}
where the Einstein tensor including the cosmological constant is
given by
\begin{equation}
G_{mn}=R_{3mn}-\frac{R_3}{2}g_{mn} -\frac{1}{l^2}g_{mn},
\end{equation}
and the Cotton tensor is
\begin{equation}
C_{mn}= \varepsilon_m~^{pq}\nabla_p
\Big(R_{3qn}-\frac{1}{4}g_{qn}R_3\Big).
\end{equation}
We note that the Cotton tensor $C_{mn}$ vanishes for
any solution to $G_{mn}=0$ for Einstein gravity, so all solutions to general
relativity are also solutions of the TMG$_\Lambda$.
Hence, the BTZ black hole solution~\cite{BTZ} appears  for the $K=0$ case.
For the $K\not=0$ case,  the warped  black hole solution for
$\nu^2=(\frac{l}{3K})^2>1$, which is asymptotic to the warped
AdS$_3$, was found  as \cite{StromingerWp,CD,OK}
\begin{equation}
ds^2_{wBH}=-\tilde{N}^2dt^2+\frac{l^4}{4\tilde{R}^2\tilde{N}^2}dr^2+l^2\tilde{R}^2\Big(d\theta+\tilde{N}^\theta
dt\Big)^2,
\end{equation}
where
\begin{eqnarray}
 && \tilde{R}^2(r)=\frac{r}{4}\left[3(\nu^2-1)r +(\nu^2+3)(r_++r_-)-4\nu\sqrt{r_+r_-(\nu^2+3)}\right],\nonumber\\
 && \tilde{N}^2(r)=\frac{l^2(\nu^2+3)(r-r_+)(r-r_-)}{4\tilde{R}^2(r)},\nonumber\\
 && \tilde{N}^\theta(r)=\frac{2\nu r-\sqrt{r_+r_-(\nu^2+3)}}{2\tilde{R}^2(r)}.
\end{eqnarray}
We note that this metric reduces to the BTZ metric in a rotating
frame when choosing $\nu^2=1$.

\section{AO type reduction of TMG$_\Lambda$}
Let us perform AO type of dimensional reduction using
the metric~\cite{AO}
\begin{equation} \label{AOmetric}
ds^2_{AO}=g_{\mu\nu}(x)dx^\mu dx^\nu + \phi^2(x)
\Big[d\theta+A_\mu (x)dx^\mu\Big]^2
\end{equation}
without conformal transformation~\cite{mkp1}. Here $\theta$ is a
coordinate that parameterizes an $S$ with period $2 \pi l$.
Hence, its isometry is factorized as ${\cal G}\times U(1)$. After
the $\theta$-integration, the action (\ref{tmg}) reduces to an
interesting two-dimensional TMG$_\Lambda$ action
\begin{eqnarray} \label{aoaction}
 {\cal I}_{\rm AOTMG_\Lambda}&=& \frac{l}{8G_3}\int d^2x \sqrt{-g} \phi
            \left(R+\frac{2}{l^2}-\frac{1}{4}\phi^2 F_{\mu\nu}F^{\mu\nu}\right)\nonumber\\
 &+& \frac{Kl}{32G_3} \int d^2x~ \phi^2\left(R\epsilon^{\mu\nu}F_{\mu\nu}+
     \phi^2\epsilon^{\mu\nu}F_{\mu\rho}F^{\rho\sigma}F_{\sigma\nu}\right).
\end{eqnarray}
Here $R$ is the 2D Ricci scalar with $R_{\mu\nu}=Rg_{\mu\nu}/2$
and $\phi$ is the dilaton. Also
$F_{\mu\nu}=\partial_{\mu}A_{\nu}-\partial_{\nu}A_{\mu}$ and
$\epsilon^{01}=1$. The Greek indices of $\mu,\nu, \rho, \cdots$
represent two dimensional tensors. Hereafter we choose $G_3/l=1/8$
for simplicity. We mention that this action was recently  used to
find AdS$_2$ solutions~\cite{ads22}.

Introducing  a dual scalar $F$ of  the Maxwell field through
\begin{equation}
F=-\frac{\epsilon^{\mu\nu}F_{\mu\nu}}{2\sqrt{-g}},\end{equation}
equations of motion for $\phi$ and $A_\mu$ are given by
\begin{eqnarray}
 && \label{EOM-phi}
    R+\frac{2}{l^2}+\frac{3}{2}\phi^2 F^2-K\phi F(R+2\phi^2F^2)=0,  \\
 && \label{EOM-A}
    \epsilon^{\mu\nu}\partial_\mu \left[\phi^3 F-\frac{K}{2}\phi^2(R+3\phi^2F^2)\right]=0,
\end{eqnarray}
respectively. The equation of motion for $g^{\mu\nu}$ is described
as follows:
\begin{eqnarray}
 && \label{EOM-g}
    g_{\mu\nu}\left(\nabla^2\phi-\frac{1}{l^2}\phi+\frac{1}{4}\phi^3F^2\right)
    -\nabla_\mu\nabla_\nu\phi\nonumber\\
 && -\frac{K}{2}\left[g_{\mu\nu}\left(\frac{1}{2}R\phi^2F+\phi^4F^3+\nabla^2(\phi^2F)\right)
                      -\nabla_\mu\nabla_\nu(\phi^2F)\right]=0.
\end{eqnarray}
Moreover, the above equation  can be further separated into the
trace part
\begin{equation}
  \label{EOM-trg}
  \nabla^2\phi-\frac{2}{l^2}\phi+\frac{1}{2}\phi^3F^2
  -K\left[\frac{1}{2}R\phi^2F+\phi^4F^3+\frac{1}{2}\nabla^2(\phi^2F)\right]=0,
\end{equation}
and  the traceless part
\begin{equation}
  \label{EOM-trlessg}
  \frac{1}{2}g_{\mu\nu}\nabla^2\phi-\nabla_\mu\nabla_\nu\phi
  -\frac{K}{2}\left[\frac{1}{2}g_{\mu\nu}\nabla^2(\phi^2F)-\nabla_\mu\nabla_\nu(\phi^2F)\right]=0.
\end{equation}
Note that the ``traceless part" does not play no role in obtaining
vacuum solutions. However, when carrying  out the perturbation
analysis in the next section, this equation imposes a nontrivial
constraint on the propagation of physical modes. This  contrasts
to our previous KK type reduction case~\cite{mkp1}, where the
``traceless part" did not put  any additional constraint on the
propagation degrees of freedom. (See the footnote 2.)

Now, we wish to find the AdS$_2$ background as a vacuum solution
to  equations of motion. In case of a constant dilaton, from Eqs.
(\ref{EOM-phi}) and (\ref{EOM-trg}), we have the condition
\begin{equation}
 (1-\frac{3}{2}K\phi F)\left(\frac{4}{l^2}-\phi^2F^2\right)=0,
\end{equation}
which implies three different relations between $\phi$ and $F$
\begin{equation}
\label{ads}
 \phi_{\pm} = \pm\frac{2}{lF},~~ \phi_{w} = \frac{2}{3KF}.
\end{equation}
Note here that for $K=l/3$, $\phi_w$ reduces to  $\phi_+$.
Assuming the line element preserving ${\cal G}=SL(2,R)$ isometry
\begin{equation} \label{ads1}
ds^2_{AdS_2}=v\left(-r^2dt^2+\frac{dr^2}{r^2}\right),
\end{equation}
we have the AdS$_2$-spacetimes, which satisfy
\begin{equation} \label{ads2}
\bar{R}=-\frac{2}{v},
~~~\bar{\phi}=u,~~~\bar{F}=e/v,
\end{equation}
where $\bar{F}_{10}=\partial_1 \bar{A}_{0}-\partial_0 \bar{A}_{1} = e$
with $\bar{A}_{0}=er$ and $\bar{A}_{1}=0$.
In order to find the whole solutions of AdS$_2$ type, we may use
the entropy function formalism~\cite{SSen} because it provides an
efficient way to obtain AdS$_2$ solution as well as entropy of
 extremal black hole~\cite{AFM,mkp1}.  The entropy
function is defined as
\begin{eqnarray} \label{entfn}
{\cal E}(u,v,e,q)=2\pi\Big[qe-{\cal F}(u,v,e)\Big],
\end{eqnarray}
where ${\cal F}(u,v,e)$ is the Lagrangian density ${\cal L}_{\rm
AOTMG_\Lambda}$ evaluated when using Eq. (\ref{ads2}),
\begin{equation}
 {\cal F}(u,v,e)=-2u+\frac{2uv}{l^2}+\frac{u^3 e^2}{2v}
   + K\frac{u^2 e}{v}\left(1-\frac{u^2 e^2}{2v}\right).
\end{equation}
Here we have equations of motion upon the variation of ${\cal E}$
with respect to $u$, $v$, and $e$
\begin{eqnarray}
 \label{fequ}&& -2+\frac{2v}{l^2}+\frac{3u^2 e^2}{2v} + K\frac{2u e}{v}\left(1-\frac{u^2 e^2}{v}\right) =0,\\
 \label{feqv}&& \frac{2u}{l^2}-\frac{u^3 e^2}{2v^2}
    -K\frac{u^2 e}{v^2}\left(1-\frac{u^2 e^2}{v}\right)=0,\\
\label{feqe} && q-\frac{u^3e}{v}-K\frac{u^2}{v}\left(1-\frac{3u^2
e^2}{2v}\right)=0,
\end{eqnarray}
respectively. Associated with the constant dilaton solutions
(\ref{ads}), we also obtain three kinds of AdS$_2$
solutions similar to the KK case as fellows.

\noindent(1) For $u=\frac{2v}{le}~(\phi_+=\frac{2}{lF})$, one has
AdS$_2$-solution with a positive charge $q$
\begin{equation}\label{set1}
u=\sqrt{\frac{ql^2}{2(l-K)}},~~~v=\frac{l^2}{4},~~~e=\sqrt{\frac{l-K}{2q}}.
\end{equation}
(2) For $u=-\frac{2v}{le}~(\phi_-=-\frac{2}{lF})$, one has
AdS$_2$-solution with a negative charge $q$
\begin{equation}\label{set2}
u=-\sqrt{\frac{-ql^2}{2(l+K)}},~~~v=\frac{l^2}{4},~~~e=-\sqrt{\frac{l+K}{-2q}}.
\end{equation}
(3) For $u=\frac{6Kl^2}{(l^2+27K^2)e}~(\phi_w=\frac{2}{3KF})$, one
has warped AdS$_2$-solution with a positive charge $q$
\begin{equation}\label{set3}
u=\sqrt{\frac{9Kql^2}{l^2+27K^2}},~~~v=\frac{9K^2l^2}{l^2+27K^2},~~~e=\sqrt{\frac{4Kl^2}{q(l^2+27K^2)}}.
\end{equation}
Then, the corresponding entropies of the extremal black holes are
given by
\begin{eqnarray}
 && S_+=  \frac{2\pi}{e}(l-K) \sim 2\pi\sqrt{\frac{ql}{6}\times \frac{12(l-K)}{l}},~~~l\ge K\\
 && S_-= \frac{2\pi}{e}(l+K) \sim 2\pi\sqrt{-\frac{ql}{6}\times \frac{12(l+K)}{l}},~~~l\ge -K\\
 && S_w= \frac{2\pi}{e} \frac{8Kl^2}{l^2+27K^2}\sim 2\pi\sqrt{\frac{ql}{6}\times
 \frac{96Kl}{l^2+27K^2}},~~~K>0.
\end{eqnarray}
The last relations ($\sim$) may  be confirmed by the Cardy
formula if $ql$ is the eigenvalue of $L_0$-operator of dual
CFT$_2$. However, the AdS/CFT correspondence is not still
specified, because we have two AdS$_2$
solutions~\cite{ads21,ads23,ads24,ads22}: one is AdS$_2$ with constant
dilaton and near-horizon chiral CFT$_2$ (AdS$_2$/CFT$_2$
correspondence, in this work), and the other is  AdS$_2$ with linear dilaton and
asymptotic CFT$_1$ (AdS$_2$/CFT$_1$ correspondence). Moreover, we note the entropy relation of
$S_+=\frac{4 \pi l}{3e}=S_w$ for the $K=l/3$ case, which implies a
close connection between solution (1) and (3).

We note that for (\ref{set1}), (\ref{set2}), and (\ref{set3}),
their background metric (\ref{AOmetric}) can be rewritten as the
extremal (warped) black holes using  Poincare coordinates
($t, r, z$), respectively~\cite{AFM}
\begin{eqnarray}
\label{3d0} ds^2_{AO}&=& \bar{g}_{\mu\nu}dx^\mu dx^\nu+(\bar{\phi})^2(d\theta+\bar{A}_\mu dx^\mu)^2 ~~\to \\
\label{3d1}  ds^2_+ &=& \frac{l^2}{4}\Bigg[-r^2dt^2+\frac{dr^2}{r^2}+(dz+rdt)^2\Bigg], \\
\label{3d2}  ds^2_- &=& \frac{l^2}{4}\Bigg[-r^2dt^2+\frac{dr^2}{r^2}+(dz-rdt)^2\Bigg], \\
 \label{3d3} ds^2_w &=& \frac{9K^2l^2}{27K^2+l^2}\Bigg[-r^2dr^2+\frac{dr^2}{r^2}+\frac{4l^2}{27K^2+l^2}(dz+rdt)^2
\Bigg]
\end{eqnarray}
with $\theta=ez$. This  show clearly  that the isometry of
$SL(2,R)\times U(1)$ also persists in the AO type of
dimensional reduction, similar to the KK case.

\section{Perturbation of AOTMG$_{\Lambda}$ on AdS$_2$}

Now, we consider the perturbation modes of the dilaton, graviton,
and dual scalar of the Maxwell field  around the AdS$_2$
background as~\cite{mkp1}
\begin{equation}
 \phi=\bar{\phi}+\varphi, ~~
 g_{\mu\nu}=\bar{g}_{\mu\nu}+h_{\mu\nu},~~
 F=\bar{F}(1+\delta F),
\end{equation}
where the bar variables denote the AdS$_2$ background as
$\bar{\phi}=u$, $\bar{g}_{\mu\nu}=v~{\rm diag}(-r^2,r^{-2})$, and
$\bar{F}=e/v$. This background corresponds definitely  to the
near-horizon geometry of the extremal BTZ (warped) black holes,
factorized as AdS$_2 \times S^1$. Here we note that the
perturbation of the Maxwell field is defined as
\begin{equation} F_{10}=\bar{F}_{10}+\delta F_{10},\end{equation}
 where $\delta
F_{10}=\partial_1 a_0 -\partial_0 a_1$ and $\delta F_{10}=-f$.
On the other hand, the perturbation fields are chosen to be~\cite{LMKim}
\begin{equation}
 h_{\mu\nu}=-h\bar{g}_{\mu\nu},~~~ \delta F=\left(h-\frac{f}{e}\right).
\end{equation}

\subsection{Perturbation with $K=0$}

First, let us briefly summarized the $K=0$ case, which is actually
equivalent to the KK reduction, because this provides a reference
case. However, this case is equivalent to the KK reduction case even
though their perturbed equations of (\ref{EOM-phi}), (\ref{EOM-A}),
(\ref{EOM-trg}), and (\ref{EOM-trlessg}) take slightly different
forms  as
 \begin{eqnarray}
 && \label{pEOM-phi}
    \delta R(h) +\frac{3u^2e^2}{v^2}\left(\frac{1}{u}\varphi+\delta F\right)=0,\\
 && \label{pEOM-A}
    u^3 \Big(\frac{3}{u}\varphi+\delta F\Big)=0,\\
 && \label{pEOM-trg}
    \bar{\nabla}^2\varphi-\frac{2\varphi}{l^2}+\frac{u^3e^2}{v^2}\left(\frac{3\varphi}{2u}+\delta F\right)=0,\\
 && \label{pEOM-trlessg}
    \left(\frac{1}{2}\bar{g}_{\mu\nu}\bar{\nabla}^2-\bar{\nabla}_\mu\bar{\nabla}_\nu\right)\varphi=0
\end{eqnarray}
with the linearized Ricci scalar $\delta R(h)=(\bar{\nabla}^2-2/v)h$.
It is known  from the counting of DOF  that all of these modes
belong to pure gauge because there is no  DOF for the
graviton $h_{mn}$ propagating on the AdS$_3$ background in the 3D
Einstein gravity.
 Therefore, it is necessary to show  that all modes of
$\varphi,h,$ and $f$ are non-propagating on the AdS$_2$ background.
 For this purpose, we compute
the on-shell exchange amplitude   by
plugging external sources $T,J_\varphi,$ and $J_f$ into Eqs.
(\ref{pEOM-phi}), (\ref{pEOM-A}) and (\ref{pEOM-trg}) without
constraint.
Under the source condition of $T=\frac{e^2}{v^2}J_f$, the on-shell exchange
amplitude take a contact form
\begin{equation}
\label{cont-form}
   \bar{A}^{K=0}=\frac{1}{2}\int d^2p
      \left[\frac{1}{u^5}J^2_f \right].
\end{equation}
As a result, the effective 2D gravity theory, which is correctly
matched with the original 3D Einstein gravity, has no physically
propagating modes.

\subsection{Perturbation with $K \neq 0$}

For the $K\neq0$ case,  perturbed equations of motion  are
complicated to be
\begin{eqnarray}
 && \label{pEOMK-phi}
    \delta R +\frac{3u^2e^2}{v^2}\left(\frac{\varphi}{u}+\delta F\right)
    -\frac{Kue}{v}\left[\delta R+\left(\frac{6u^2e^2}{v^2}-\frac{2}{v}\right)\left(\frac{\varphi}{u}+\delta F\right)\right]=0,\\
 && \label{pEOMK-A}
    u^3\left(\frac{3\varphi}{u}+\delta F\right)
    -\frac{Ku^2 v}{2e}\left[\delta R+\left(\frac{12u^2e^2}{v^2}-\frac{4}{v}\right)\frac{\varphi}{u}+\frac{6u^2e^2}{v^2}\delta F\right]=0,\\
 && \label{pEOMK-trg}
    \bar{\nabla}^2\varphi-\frac{2\varphi}{l^2}+\frac{u^3e^2}{v^2}\left(\frac{3\varphi}{2u}+\delta F\right)
   \nonumber\\
 && -\frac{Ku^2e}{2v}\left[\delta R+\left(\bar{\nabla}^2-\frac{2}{v}\right)\left(\frac{2\varphi}{u}+\delta F\right)
   +\frac{2ue^2}{v^2}\varphi +\frac{6u^2e^2}{v^2}\left(\frac{\varphi}{u}+\delta
   F\right)\right]=0,\\
 && \label{pEOMK-trlessg}
   \left(\frac{1}{2}\bar{g}_{\mu\nu}\bar{\nabla}^2-\bar{\nabla}_\mu\bar{\nabla}_\nu\right)
   \left[\varphi-\frac{Ku^2e}{2v}\left(\frac{2\varphi}{u}+\delta
   F\right)\right]=0.
\end{eqnarray}
Let us first consider the two AdS$_2$ solutions (\ref{set1}) and
(\ref{set2}). Considering  the relation $v=u^2e^2$, and inserting $\delta R$ in Eq.
(\ref{pEOMK-A}) into Eq. (\ref{pEOMK-phi}), we obtain a relation
between $\delta F$ and $\varphi$ as
\begin{equation}
\label{Fphi}
 \delta F = \alpha_\pm\left(\frac{\varphi}{u}\right),
\end{equation}
where $\alpha_\pm$ is determined to be
\begin{equation}\label{alpha}
 \alpha_\pm = - \left(\frac{8K \mp 3l}{4K \mp l}\right).
\end{equation}
 On the other hand, $\delta R$ in
either Eq. (\ref{pEOMK-phi}) or Eq. (\ref{pEOMK-A}) can be
simplified  to show the relation
\begin{equation}
 \label{Rphi}
 \delta R= \pm\frac{16e}{l^3} \left(\frac{8K \mp 3l}{4K \mp l}\right)
 \varphi.
\end{equation}
At this point, we check that for $K=0$ case, $\alpha_\pm=-3$,
which gives $\delta F=-3\varphi/u$.  In this case, one finds that $\delta R=(6/uv)\varphi$. It
again returns to Eq. (\ref{pEOM-phi}), implying an unusual
propagation of $(\bar{\nabla}^2-2/v)^2h=0$. Considering  Eq.
(\ref{pEOMK-trlessg}) leads to  the constraint\footnote{At this
stage, it seems appropriate to comment on the case of the
KK reduction \cite{mkp1}. In this case, the linearized equation of
the ``traceless part" is given by
\begin{equation}
\label{pEOMK-trlessg0}
   \left(\frac{1}{2}\bar{g}_{\mu\nu}\bar{\nabla}^2-\bar{\nabla}_\mu\bar{\nabla}_\nu\right)
   \left[\varphi-\frac{Ke}{2v} \delta F \right]=0.
\end{equation}
In contrast to  the AO type reduction, this equation
does not impose any additional constraint  because
$\varphi=\frac{Ke}{2v} \delta F$ is the same mode relation
appeared in Eq. (4.17) in Ref.~\cite{mkp1}.} on the mode relation
\begin{equation} \label{constraint-eq}
\frac{\varphi}{u}=\frac{1}{2}\Big[\frac{Kue}{v-Kue} \Big]\delta F.
\end{equation}
With $u=\pm \frac{2v}{le}$, the above constraint  leads to a
relation between $\varphi$ and $\delta F$  as
\begin{equation}\label{ads2trlessg}
\frac{\varphi}{u}=\Big[\frac{K}{\pm l-2K} \Big]\delta F,
\end{equation}
respectively.

It is now important to note that  for compatibility,  Eqs.
(\ref{Fphi}) and (\ref{ads2trlessg}) leads to the condition
\begin{equation} \label {k-cond} K=\pm \frac{l}{3}\end{equation} for the two AdS$_2$
solutions of $u=\pm \frac{2v}{le}$, respectively. This
compatibility condition could also  be derived  by eliminating
$\varphi$ in Eqs. (\ref{pEOMK-phi}) and (\ref{pEOMK-A}) when using
(\ref{ads2trlessg}).  We rewrite them as a matrix equation
\begin{eqnarray} \label{mateq}
 \left(\begin{array}{cc}
 \frac{Kl}{4} & \frac{(4K\mp l)(K\mp l)}{(2K\mp l)l}\\
 \frac{2K\mp l}{l} & \frac{4(8K\mp 3l)(K\mp l)}{(2K\mp l)l^3}
 \end{array} \right)
 \left(\begin{array}{c}\delta R \\ \delta F\end{array}\right)
 =
 \left(\begin{array}{c}0\\0\end{array}\right) \to M_\pm {\cal H}=0
\end{eqnarray}
with ${\cal H}=(\delta R, \delta F)$.  Then, $\delta R$ and $\delta
F$ have nontrivial solutions iff the determinant of $M_\pm$ is
zero as
\begin{equation}
{\rm det}M_\pm=0 \to \pm\frac{(3K \mp l)(K\mp l)}{(2K\mp l)l}=0
\end{equation}
 which  implies  Eq. (\ref{k-cond}).  Thus,  the
 linearized equation of the traceless part puts on an important constraint such that
\begin{equation}
\frac{\varphi}{u}=\delta F. \end{equation}

Considering the compatibility condition of $K= \pm l/3$, we
 obtain the  relation between the perturbed
fields as
\begin{eqnarray}
 \left(\bar{\nabla}^2-\frac{2}{v}\right)h=-\frac{2}{u v} \varphi = -\frac{2}{v}\delta F.
\end{eqnarray}
On the other hand,  the remaining equation (\ref{pEOMK-trg}) takes
the form
\begin{equation}
\left(1\mp\frac{3K}{l}\right)\left(\bar{\nabla}^2+\frac{2}{v}\right)\varphi=0,
\end{equation}
which does not provide  any useful  information  when  imposing
the condition of $K=\pm l/3$.

At this stage, we note that  the previous approach of linearized equations  without
sources did not show clearly what kind of modes are really
propagating  on the AdS$_2$ background.  Hence, we need to take into account
the on-shell exchange amplitude with external sources. We note
that in  Appendix, we initially compute the on-shell exchange
amplitude without the constraint. Then, we  require the condition
(\ref{k-cond}) on   the on-shell exchange amplitude.

For the two AdS$_2$ solutions, the coupled equations in
(\ref{mateq}) take the form with external sources
\begin{eqnarray}
\left(\begin{array}{cc}
 \frac{Kl}{4} & \frac{(4K\mp l)(K\mp l)}{(2K\mp l)l}\\
 \frac{2K\mp l}{l} & \frac{4(8K\mp 3l)(K\mp l)}{(2K\mp l)l^3}
 \end{array} \right)
 \left(\begin{array}{c}\delta R \\ \delta F\end{array}\right)
 =
 \left(\begin{array}{c}\pm\frac{1}{u^3}J_f\\\mp J_\varphi\end{array}\right)
\end{eqnarray}
which allow  $\delta R$ and  $\delta F$ to  express in terms of
the sources as
\begin{eqnarray}
 \label{so1} && \delta R = \frac{32e^3(8K\mp 3l)J_f-(4K\mp l)l^5 J_\varphi}{(3K\mp l)l^5},\\
 \label{so2} && \delta F= \frac{(2K\mp l)[ Kl^5 J_\varphi -32e^3(2K\mp l)J_f]}{4l^3(3K\mp l)(K\mp l)}.
\end{eqnarray}
On the other hand, using Eqs. (\ref{so1}) and (\ref{so2}), the
remaining equation (\ref{pEOMK-g1s}) (equivalently,
(\ref{pEOMK-trg}) with $-T$) leads to the source condition
\begin{equation} \label{soscod}
T=\frac{e^2}{v^2}J_f.
\end{equation}
Finally, making use of (\ref{soscod}), (\ref{so1}),
(\ref{so2}), and (\ref{constraint-eq}), the Fourier-transformed
on-shell amplitude
\begin{equation}
\label{defSaction}
 \bar{A}^K_\pm=\frac{1}{2}\int d^2p~ \left[v \varphi(p) J_\varphi
          +v h(p) \left(-T+\frac{e^2}{v^2}J_f \right)-\frac{e^2}{v}\delta F(p)
 J_f\right]
\end{equation}
reduces to
\begin{equation} \label{defSactionf}
 \bar{A}^K_\pm=\frac{1}{u^5l}\int d^2p~
 \frac{1 }{{\rm det} M_\pm}\Bigg[(2K\mp l)J^2_f-\frac{K^2u^6v^2}{2K\mp l}J^2_\varphi
 \Bigg],
\end{equation}
which  contains contact terms without poles. Hence, this  does not
contribute to the interaction between the separate sources. This
is also recovered from  Eq. (\ref{nonkamp}) with the source
condition (\ref{soscod}). However, we observe that Eq.
(\ref{defSactionf}) blows up because $ {\rm det} M_\pm=0$ when
imposing the compatibility condition $K=\pm l/3$. Hence, the
AO type reduction is not a proper way to obtain  propagating
modes on the near-horizon geometry of AdS$_2 \times S^2$ of
extremal black holes.

Now, let us consider the perturbation around the warped AdS$_2$
solution (\ref{set3}). By making use of $u=2v/3Ke$,
(\ref{pEOMK-phi}) and (\ref{pEOMK-A}) give
\begin{equation}\label{wFu}
 \delta F=-\frac{2(27K^2-2l^2)}{27K^2-5l^2}\left(\frac{\varphi}{u}\right),
\end{equation}
while Eq. (\ref{pEOMK-trlessg}) implies
\begin{equation}\label{warptrlessg}
\delta F=\frac{\varphi}{u}.
\end{equation}
Thus, equating (\ref{wFu}) and (\ref{warptrlessg})  leads again to
the compatibility condition of $K=\pm l/3$ even for the warped
AdS$_2$ solution. Moreover,  Eqs. (\ref{pEOMK-phi}) and
(\ref{pEOMK-A}) can be rewritten as
\begin{eqnarray} \label{mateq-wp}
 \left(\begin{array}{cc}
 1 & \frac{20}{9K^2}-\frac{12}{l^2}\\
 1 & -\frac{16}{9K^2}+\frac{24}{l^2}
 \end{array} \right)
 \left(\begin{array}{c}\delta R \\ \delta F\end{array}\right)
 =
 \left(\begin{array}{c}0\\0\end{array}\right) \to M_w {\cal H}=0.
\end{eqnarray}
From the condition of  nontrivial solutions for $\delta R$ and $\delta F$ as
\begin{equation}
{\rm det}M_w=\frac{4(9K^2-l^2)}{K^2l^2}=0,
\end{equation}
we again confirm the compatibility condition $K=\pm l/3$ for the warped
AdS$_2$ solution. The perturbed fields are satisfied as
\begin{equation}
 \left(\bar{\nabla}^2-\frac{2}{v}\right)h= -\frac{8}{l^2}\delta F
 = -\frac{8}{u l^2}\varphi
\end{equation}
while the remaining equation (\ref{pEOMK-trg})  does not give rise to any further information
as before, upon using the compatibility condition.

On the other hand, for the warped AdS$_2$ solution, the coupled equations
(\ref{mateq-wp}) take the form with the external sources
\begin{eqnarray}
 \left(\begin{array}{cc}
 1 & \frac{20}{9K^2}-\frac{12}{l^2}\\
 1 & -\frac{16}{9K^2}+\frac{24}{l^2}
 \end{array} \right)
 \left(\begin{array}{c}\delta R \\ \delta F\end{array}\right)
 =
 \left(\begin{array}{c}\frac{(27K^2+l^2)^3e^3}{162K^5l^6}J_f\\ -3
 J_\varphi\end{array}\right).
\end{eqnarray}
Now, $\delta R$ and  $\delta F$ are
rewritten  in terms of the external sources as
\begin{eqnarray}\label{so1-w}
 \delta R &=& \frac{1}{(9K^2-l^2)}\left(
             \frac{(27K^2+l^2)^3(27K^2-2l^2)e^3}{729K^5l^6}J_f
             - \frac{1}{3}(27K^2-5l^2)J_\varphi\right),\\
 \label{so2-w} \delta F &=& \frac{-1}{(9K^2-l^2)}\left(
             \frac{(27K^2+l^2)^3e^3}{648K^3l^4(9K^2-l^2)}J_f
             +\frac{3K^2l^2}{4(9K^2-l^2)}J_\varphi
             \right).
\end{eqnarray}
On the other hand, using  (\ref{so1-w}) and (\ref{so2-w}), the
remaining equation (\ref{pEOMK-g1s}) (equivalently,
(\ref{pEOMK-trg}) with $-T$) leads to the source condition (\ref{soscod}).

Finally, using  (\ref{so1-w}) and  (\ref{so2-w}), one
can obtain the Fourier-transformed on-shell amplitude for the
warped AdS$_2$ case as
\begin{equation} \label{defSactionf-w}
 \bar{A}^K_w=\frac{2}{K^2l^2}\int d^2p~
  \frac{1 }{{\rm det} M_w}\Bigg[\frac{(27K^2+l^2)^4e^5}{2^33^6K^5l^6}J^2_f
 -\frac{3^4K^5l^6}{2e(27K^2+l^2)^2}J^2_\varphi
 \Bigg],
\end{equation}
which is exactly the same with Eq. (\ref{Amwarp}) when imposing
the source condition.

\section{Discussions}

We have studied the topologically massive gravity with a negative
cosmological constant on AdS$_2$ spacetimes by making use of the
AO type of  dimensional reduction. We have obtained
that for a constant dilaton, the two-dimensional model of
AOTMG$_{\Lambda}$ admits three AdS$_2$ solutions. As was shown in
Eqs. (\ref{3d1}), (\ref{3d2}), and  (\ref{3d3}), these are related
to AdS$_3$ with positive/negative charge and warped
AdS$_3$-solution with an identification upon uplifting three
dimensions similar to the KK  reduction case. However,
it turns out that there is no propagating massive mode on AdS$_2$,
in contrast to the KK  case. This shows that the
AOTMG$_\Lambda$ based on the AO  type of the
dimensional reduction  is not appropriate for describing a massive
propagation of the TMG$_\Lambda$ on AdS$_2$ spacetimes, even
though it was successfully used to derive the entropies of
extremal BTZ and warped black holes when applying the entropy
function formalism.

How we could understand  the disappearance of massive modes under
the AO type of  dimensional reduction? Is it related to the isometry?
 For
topologically massive gravity without a negative cosmological
constant, it was known that the exact theory has no nontrivial
solutions that admits a hypersurface-orthogonal Killing
vector~\cite{AN}. This means that assuming too much isometry may
eliminate all of  propagating modes. In this work, the isometry of
AdS$_3$, $SL(2,R)\times\overline{SL(2,R)}$ is broken to the
isometry of AdS$_2$, $SL(2,R)\times U(1)$ when performing the
AO reduction. At this stage, we note that two types
of dimensional reduction provide the same isometry of
$SL(2,R)\times U(1)$.  Also these give the same field contents of
$\varphi,\delta F,$ and $ h$. Therefore, it is reasonable to consider that the dimensional reduction
does not eliminate  all propagating modes.

An important thing to remark is not an isometry breaking, but
the role of the dilaton.  It seems that replacing  $g_{\mu\nu}$ by
$ \phi^2g_{\mu\nu}$ in Eq. (\ref{AOmetric}) leads to  the metric
for the KK reduction. However, we emphasize  that  this
replacement  does not lead to the 2DTMG$_\Lambda$ action obtained by the KK reduction:
\begin{eqnarray}
 {\cal I}_{\rm 2DTMG_\Lambda}&=& \frac{l}{8G_3}\int d^2x \sqrt{-g}
            (\phi R+ \frac{2}{\phi}(\nabla\phi)^2+\frac{2}{l^2}\phi^3-\frac{\phi}{4} F_{\mu\nu}F^{\mu\nu})\nonumber \\
 &+& \frac{Kl}{32G_3} \int d^2x~(R\epsilon^{\mu\nu}F_{\mu\nu}+
     \epsilon^{\mu\nu}F_{\mu\rho}F^{\rho\sigma}F_{\sigma\nu}),
\end{eqnarray}
where  the kinetic term of $\phi$ appears in the Einstein action,
while the dilaton never appears in the Chern-Simons terms.
Actually, this operation leads to a different form of
\begin{equation}
 {\cal I}_{\rm 2DTMG_\Lambda}+\frac{Kl}{32G_3}\int d^2x~
       \frac{2}{\phi}\nabla_\mu\phi\nabla^\mu(\epsilon^{\mu\nu}F_{\mu\nu}).
\end{equation}
For $K=0$, we have proven that two are equivalent~\cite{Cad}
because two of ${\cal I}_{\rm 2DTMG_\Lambda}$ and ${\cal I}_{\rm
AOTMG_\Lambda}$ describe the same on-shell amplitude
(\ref{cont-form}). However, for $K\not=0$, two are inequivalent
because ${\cal I}_{\rm 2DTMG_\Lambda}$ ``does not" contain the
dilaton in the Chern-Simons terms, whereas  ${\cal I}_{\rm
AOTMG_\Lambda}$ ``does" contain the dilaton in the Chern-Simons
terms. In the KK reduction case, the disappearance of the dilaton
persists in the linearized perturbation theory. Hence, there is obviously no
coupling between $\varphi$ and $\delta F$. This makes $\delta F$
massive for the $K\not=0$ case.    On the other hand, as ${\cal
I}_{\rm AOTMG_\Lambda}$ is shown in Eq. (\ref{aoaction}), the
dilaton appears in the Chern-Simons terms. As a result, all perturbed
modes of $\varphi,\delta F$ and $h$ become coupled to each other,
eliminating a massive mode in contrast to the KK reduction case.

At this stage, we note that  the constraint (\ref{constraint-eq})
(compatibility condition (\ref{k-cond})) obtained from the
traceless part is not compatible with the finiteness of on-shell
amplitude for two AdS$_2$ solutions because it blows up as
(\ref{defSactionf}) shows. Also, as is shown in
(\ref{defSactionf-w}), the  on-shell amplitude blows up  for the
warped solution. This implies that the 2D action $ {\cal I}_{\rm
AOTMG_\Lambda}$ is not suitable for describing the near-horizon
geometry AdS$_2 \times S^1$ of extremal black holes.

In conclusion, the AO type of dimensional reduction
with ``a constant dilaton"  is not a correct way to study the
near-horizon geometry AdS$_2 \times S^1$ of extremal black hole
from topologically massive gravity, while the KK type of
dimensional reduction provides a promising scheme to investigate
the near-horizon geometry of extremal black
hole ~\cite{mkp1}.  Importantly, we have observed  that  in the
presence of the gravitational Chern-Simons terms (for $K\not=0$),
two actions are inequivalent and thus,  the non-equivalence of
their on-shell amplitudes is obtained.

 Finally, we would like to mention that  the AO type of
dimensional reduction is useful for studying the AdS$_2$ black
hole without the gravitational Chern-Simons terms~\cite{Nav}.

\section*{Acknowledgement}

The authors thank D. Grumiller for communication on physical
degrees of freedom. Y.-W. Kim was supported by the Korea Research
Foundation Grant funded by Korea Government (MOEHRD):
KRF-2007-359-C00007. Two of us (Y. S. Myung and Y.-J. Park) were
supported by the National Research Foundation of Korea (NRF) grant
funded by the Korea government (MEST) through the Center for
Quantum Spacetime (CQUeST) of Sogang University with grant number
2005-0049409.

\section*{Appendix: On-shell exchange amplitudes with external sources  ($K\neq 0$ case without constraint)}
Let us start with the bilinear action with external sources
\begin{equation}
 \label{Saction}
 A^{K}=\int d^2 x \left[\delta_2{\cal L}_{\rm
AOTMG_\Lambda}(h,\varphi,f)
 +\sqrt{-\bar{g}}\left(\varphi J_\varphi+h_{\mu\nu}T^{\mu\nu}+\frac{e}{v^2}fJ_f\right)\right].
\end{equation}
Then, the linearized equations with external sources
are given by
\begin{eqnarray}
 && \label{pEOMK-phis}
 \delta R+\frac{3u^2e^2}{v^2}\left(\frac{\varphi}{u}+\delta F\right)
 -\frac{Kue}{v}\left[\delta R+\left(\frac{6u^2e^2}{v^2}-\frac{2}{v}\right)\left(\frac{\varphi}{u}+\delta
 F\right)\right]=-J_\varphi,\\
 && \label{pEOMK-As}
 u^3\Big(\frac{3\varphi}{u}+\delta  F\Big)
 -\frac{Ku^2v}{2e}\left[\delta R+\left(\frac{12u^2e^2}{v^2}-\frac{4}{v}\right)\frac{\varphi}{u} +\frac{6u^2e^2}{v^2}\delta F\right]=-J_f, \\
 && \label{pEOMK-g1s}
 \bar{\nabla}^2\varphi-\frac{2}{l^2}\varphi+\frac{u^3e^2}{v^2}\left(\frac{3\varphi}{2u}+\delta F\right)\nonumber\\
 &&
 -\frac{Ku^2e}{2v}\left[\delta R+\left(\bar{\nabla}^2-\frac{2}{v}\right)\left(\frac{2\varphi}{u}+\delta F\right)
 +\frac{2ue^2}{v^2}\varphi+\frac{6u^2e^2}{v^2}\left(\frac{\varphi}{u}+\delta F\right)
 \right]=-T.
\end{eqnarray}
If the above sources are  turned off, these  are the same
equations of (\ref{pEOMK-phi}), (\ref{pEOMK-A}), and
(\ref{pEOMK-trg}), respectively. Hence we could follow the same
diagonalizing process in Sec. 4.2 with the sources. As was shown
before, we first consider the AdS$_2$ solutions ($\phi_\pm$) with
$u=\pm(2v/le)$. By solving Eq. (\ref{pEOMK-As}) for $\delta R$ and
inserting it into Eq. (\ref{pEOMK-phis}), we have $\delta F$ with
the sources as
\begin{equation}
 \label{delFs}
 \delta F=\alpha_\pm\frac{\varphi}{u}+\alpha_s,
\end{equation}
where $\alpha_\pm$ is given by Eq. (\ref{alpha}) and $\alpha_s$
\begin{eqnarray}
 \alpha_s=\mp\frac{\frac{Kl^3}{4}J_\varphi+\frac{8e^3}{l}\left(1\mp\frac{2K}{l}\right)J_f}{(K\mp l)(4K\mp l)}.
\end{eqnarray}
Now $\delta R$ can be read off from Eq. (\ref{pEOMK-As}) as
\begin{equation}
 \label{delRs}
 \delta R=\left(\bar{\nabla}^2-\frac{2}{v}\right)h=\beta_\pm\varphi+\beta_s,
\end{equation}
where $\beta_\pm=(16e/l^3)\alpha_\pm$  and $\beta_s$
\begin{equation}
 \beta_s=\frac{32e^3}{Kl^4}J_f
     \pm \frac{(6K\mp l)l\left[J_\varphi\mp \frac{32e^3(2K\mp l)}{Kl^5}J_f\right]}{(K\mp l)(4K\mp l)}.
\end{equation}
Finally, making use of (\ref{delFs}) and  (\ref{delRs}), we
obtain from Eq. (\ref{pEOMK-g1s}) the dilaton equation expressed in terms of sources
\begin{eqnarray}\label{delphis}
   \left(\bar{\nabla}^2-\frac{2}{v}\right)\varphi&=&-\left(\frac{4K\mp l}{3K\mp l}\right)\left(T-\frac{e^2}{v^2}J_f \right)\nonumber\\
  &+&\frac{1}{(K\mp l)(3K\mp l)}\left(\bar{\nabla}^2-\frac{2}{v}\right)\left[\frac{4Ke^2(2K\mp l)}{l^2}J_f\mp \frac{K^2l^3}{8e}J_\varphi\right].
\end{eqnarray}
After obtaining the Fourier-transformed
fluctuations from (\ref{delFs}), (\ref{delRs}),
and (\ref{delphis}), and making a tedious calculation, we arrive at the
Fourier-transformed on-shell amplitude induced by the external
sources as
\begin{eqnarray} \label{nonkamp}
 && \bar{A}^K_{\pm}=\frac{1}{2}\int d^2 p
  \left[\frac{1}{(K\mp l)(3K\mp l)}
   {\left(\pm\frac{32e^5(2K\mp l)^2}{l^5}J^2_f\mp\frac{K^2l^5}{32e}J^2_\varphi\right)}
               \right.\nonumber\\
 &&~~~~~~~~~~ \left.  \pm\frac{16e^3(8K\mp l)}{l^3(3K\mp l)}\frac{\left(T-\frac{e^2}{v^2}J_f\right)J_f}
            {\bar{p}^2+\frac{2}{v}}
            \pm\frac{4e(8K\mp l)}{l(3K\mp l)}
            \frac{\left(T-\frac{e^2}{v^2}J_f\right)^2}{\left(\bar{p}^2+\frac{2}{v}\right)^2}\right]
\end{eqnarray}
for  two $AdS_{2}$ ($\phi_\pm$) solutions. We easily check that
for the $K=0$ case with  the source condition
$T=\frac{e^2}{v^2}J_f$,
the Fourier-transformed on-shell amplitude shows no pole.

For the warped AdS$_2$ solution with $u=2v/3Ke$, by
repeating the tedious whole procedure as the AdS$_2$ case, we
finally arrive at the Fourier-transformed on-shell amplitude
\begin{eqnarray} \label{Amwarp}
 && \bar{A}^K_{w}=\frac{1}{2}\int d^2p \frac{1}{(9K^2-l^2)}
 \left[\frac{(27K^2+l^2)^4e^5}{27\cdot 6^3 K^5l^6}J^2_f
     -\frac{3^4 K^5l^6}{2e(27K^2+l^2)^2}J^2_\varphi
 \right. \nonumber\\
 &&~~~~~~~~~~~~~~~~~~~ \left.+\frac{e(27K^2+l^2)(27K^2-2l^2)}{9Kl^2} \frac{
           \left(T-\frac{e^2}{v^2}J_f\right)^2}{(\bar{p}^2+\frac{2}{v})^2}
     \right].
\end{eqnarray}
We mention that  under the source condition $T=\frac{e^2}{v^2}J_f$,
the Fourier-transformed on-shell amplitude also shows no pole.
However,  both $\bar{A}^K_{\pm}$ and $ \bar{A}^K_{w}$ blow up in the
limit of $K\rightarrow l/3$, showing a signal for the failure of the
action ${\cal L}_{\rm AOTMG_\Lambda}$, which is introduced in order
to describe a massive propagating mode in the AdS$_2$ spacetimes.

\end{document}